# Multilayer Approach to Defend Phishing Attacks

CYNTHIA DHINAKARAN[1], DHINAHARAN NAGAMALAI[2], JAE KWANG LEE[1]
[1]Department of Computer Engineering
[1]Hannam University, [2]Wireilla Net Solutions
[1]SOUTH KOREA

## Abstract

Spam messes up users inbox, consumes resources and spread attacks like DDoS, MiM, phishing etc. Phishing is a byproduct of email and causes financial loss to users and loss of reputation to financial institutions. In this paper we examine the characteristics of phishing and technology used by Phishers. In order to counter anti-phishing technology, phishers change their mode of operation; therefore a continuous evaluation of phishing only helps us combat phisher effectiveness. In our study, we collected seven hundred thousand spam from a corporate server for a period of 13 months from February 2008 to February 2009. From the collected data, we identified different kinds of phishing scams and mode of operation. Our observation shows that phishers are dynamic and depend more on social engineering techniques rather than software vulnerabilities. We believe that this study will develop more efficient anti-phishing methodologies. Based on our analysis, we developed an anti-phishing methodology and implemented in our network. The results show that this approach is highly effective to prevent phishing attacks. The proposed approach reduced more than 80% of the false negatives and more than 95% of phishing attacks in our network.

## 1 Introduction

Our modern world believes that email is great tool for the management and dissemination of information. Email provides an easy path for data flow in and out of a controlled local network. At the same time, the market of illegal business through Internet is growing. Phishing is one kind of illegal business that challenges online transaction systems. Phishing mail reaches millions of end users by spam[5] and spammers are using well designed software tools as well as social engineering methodologies to reach these end users. Phishing tools are highly sophisticated such that phishers can fix the duration of the attack, frequency of attack to the particular target, subject of the phishing mail, contents of the message, hiding the source of the mail etc., Some of the sample automated tools are Email Spoofer, Bulk Mailer, Aneima 2.0, Avalanche 3.5, Euthanasia etc[4]. Phishing is a kind of spam causing reputation, economic and man power loss. Phishers heavily utilize social engineering techniques to lure email users and divulge their valuable data. The data includes name, pass phrase, social security numbers, telephone numbers, address, and email accounts etc. Many types of phishing have been reported in recent years, including: a) the most common way of diverting the end-user to fraudulent website controlled by a phisher by clicking a hyperlink available in an email; the fraudulent website then asks the user to divulge the sensitive information and this leads to loss. b) asking the user to contact a phishers telephone number or fax number. c) through instant messaging from unknown contacts leading to password and identity theft.

Most phishing activities are associated with financial institutions, ecommerce websites, and online greeting card services. Phishing is closely related to spamming. We identify that some spammers are doing spamming as well as phishing simultaneously. Anti-spam technology ranges from simple content filters to sophisticated software based on various algorithms such as Bayesian and are widely used to stop spam. But spammers always find new ways to reach users inbox. The arms race between anti-spam software developers and spammers causes huge concern to the Internet community. Anti-spam filters are highly useful to defend the phishing scams and, apart from this, there are plenty of tools that are available to fight against phishing scams. For example CallingId, Cloudmark, Earthlink, eBay, Netcraft, Trustwatch, Spoofguard, Site Advisor [3] are some of the anti-phishing tools worth mentioning here. But unfortunately most of these tools do not provide adequate protection to the end users. According to the [3] research, the best anti-phishing software tool identifies 50% of false positives. There are more than 2 dozen free anti-phishing tools available to end users but because phishers change their tactics regularly based on anti-phishers take down approach as well as anti-phishing software, the anti-phishing tools are not able to cope with the technology of phishers. Phishing is an attack exploiting human vulnerabilities rather than technical vulnerabilities. According to the Gardener [16] survey, phishing will grow in near future since it is a profitable business. In this paper we will discuss how the phishers operate a successful businesses and the kind of technology they use and how they make use of social engineering methods. Finally we conclude with effective countermeasures to this problem. We believe that this study will develop more efficient anti-phishing methodologies.

In our study, we collected seven hundred thousand spam from a corporate server for a period of 13 months from February 2008 to February 2009. From the collected data, we identified different kinds of phishing scams and mode of operation. Our observation shows that phishers are dynamic and depend more on social engineering techniques rather than software vulnerabilities. We believe that this study will develop more efficient anti-phishing methodologies. Based on our analysis, we developed an anti-phishing methodology and implemented in our network from February 2009 to February 2010. The results show that this approach is highly effective to prevent phishing attacks. The proposed approach reduced more than 80% of the false negatives and more than 95% of phishing attacks in our network.In this paper we propose a multi layer approach to defend phishing attacks. This approach is a combination fine tuning of spamming techniques, periodically classifying false negative to spam, block the attackers IP addresses, using source and content filters to mark phishing attempt, educate and train users, report to service providers and target Institution. The results show that this approach is highly effective to prevent phishing attacks. The proposed approach reduced more than 80% of the false negatives and more than 95% of phishing attacks in our network.

The remaining sections are organized as follows. Section 2 provides background on phishing and spam and the effects. Section 3 provides data collection and experimental results. In section 4 and 5 we describe the methodologies used by phishers. Section 6 provides the countermeasure proposed by us to the phishing scam. We conclude in section 7.

## 2 Related Work

[1] Authors analyze the human factors involved in phishing attacks and suggest few techniques as remedy. According to their study, most phishing emails are not signed by a person but by position or organization like Account Manager, PayPal etc. Phishers are least bothered about the design, spelling errors and copyright information in their web site. Even legitimate web pages emphasize more on security issues, creates panic among the user community. Their study reveals that some of the third party security endorsements are not well received by the users. Also the email stimuli is more phishy than web stimuli and the phishers use automated phone messages to increase the vulnerability of the users. [2]The authors examine the characteristics of phishing URLs, domain and machines used to host phishing domains and determined phishers are more active than ordinary users and they activate their sites immediately after registering their domain and bring it down whenever desired. Phishing URLs normally contain the target's name but the length of the target name is different from the original, for phishing domain names are significantly different from regular English character frequency. Their study also reveals that most phishing hosted servers are not available in standard ports. Phishers usually do this to avoid the identification of life span of the web site. [3] Present a multi-layer approach to defend DDoS attacks caused by spam. Their study reveals the effectiveness of SURBL, DNSBLs, content filters and present characteristics of virus, worms and Trojans accompanied by spam as an attachment. The multi-layer approach defends against DDoS attacks effectively. [4] The authors observe that spammers use automated tools to send spam with attachment. The main features of this software are hiding sender's identity, randomly selecting text messages, identifying open relay machines, mass mailing capability and defining spamming duration. They table that heavy users email accounts are more vulnerable than relatively new email accounts[5][20][21]. The authors test ten well- known anti-phishing software tools to evaluate the performance of anti-phishing software tools. Their study reveals that none of the tools are 100% accurate and only one tool works better than rest with more than 42% of false positive. The authors also identify many of the phishing sites taken down within hours. Some of these tools didn't detect phishing.

In our study, we identify phishers sending mail from different sources or different IP addresses. The take down approach works good to chase phishers temporarily but not to eliminate them totally. The authors identify that the performance of anti-phishing tools vary depending on the source of phishing URLs. These tools are based on black lists, white lists, heuristics and community ratings. Most of these technologies are similar to anti-spam technologies with little modifications for phishing.

## 3 Data Collection

In our study we characterized phishing mail from a collection of over 700 000 spam over a period of 13 months from Feb 2008 to Feb 2009 from a corporate mail server. The mail server provides services to 280 users with 30 group email accounts and 290 individual mail accounts. The speed of Internet connection is 100 Mpbs with 40 Mbps upload and download speed; due to security and privacy concerns, we are unable to disclose the real domain name. In order to segregate spam from legitimate mails, we conducted standard spam detection tests on our server. The spam mails detected by these techniques were directed to the spam trap that is set up on mail server. The spam mails were usually related to finance, pharmacy, business promotion, adultery services and viruses. Considerable portion of these spam mails were used for phishing, DDoS attacks, MiM attacks etc. From this collection, we separated phishing mails by a small program written in Python. In this paper we study the characteristics of phishing and the technology used by phishers. In order to counter anti-phishing technology, phishers change their mode of operation frequently; this required us to perform a continual evaluation of the characteristics of phishing and their technology. The results of these evaluations help us to enhance existing anti-phishing technology and combat phishers effectively.

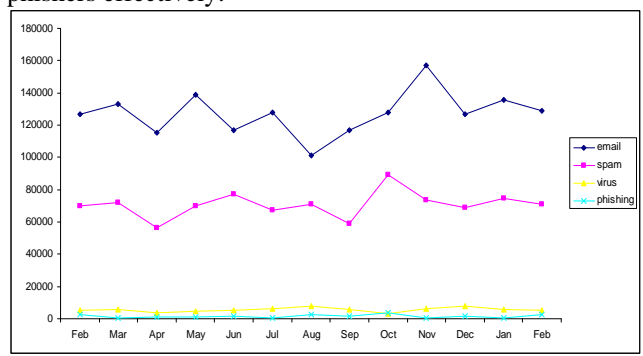

Figure 1. Mail traffic

Figure 1 shows the incoming mail traffic of our mail server for 13 months from Feb 2008 to Feb 2009. X axis identifies the month and Y axis identifies the number of mail, spam, virus, and phishing mail received by end users. As seen from the graph, the spam traffic is not related to virus and phishing quantitatively. But our analysis shows common factors in spam and phishing like senders, and the technology used. Roughly the number of phishing mails ranges from 369 to 3459 with an average rate of 1516 per month. This statistic does not include false negatives. We believe that this analysis could be useful to develop more efficient anti-phishing techniques

# 4  Phishing Methodologies

Since phishers use combination of spam and other techniques to reach end users, we have generated a table of our findings in this section.

Our spam collection contains major phishing scams as shown in Table 1. It is evident that other virus, worm, and Trojan entities are used to phish end users. There are two types of phishing methodologies detected through our data collection: one method is a simple link in the spam mail where phishers lure end users to divulge their sensitive information and the other is approaching end users with Trojan, malware & viruses. The following are the malware, Trojan and virus entities used for phishing.

| Name of virus, worm, Trojan |
| --- |
| Spam.Phish.url |
| Spam.Hoax.HOAX_PHISH_FORGED_PAYPAL |
| Clm.HTML.Phishing.Pay-110 |
| Spam.Hoax.HOAX_PHISH_FORGED_EBAY |
| Spam.Hoax.HOAX_PHISH_FORGED_CITIBNK |
| Trojan.spy.html.bankfraud.od |

Table 1:List of Virus,worm,Trojon

## 4.1 Spam.Phish.url
This is a common phishing scam used to target all the types of users with all brands. The major brand spoofed by spam.Phish.url are Citi, Paypal, 53, fifth third bank and some other small brands. The number of spam mail received by these scam are far more than others. As shown in Table 1, the number of such mails is 100 times more than any other particular phishing scam. We identified two types of phishing methods for spam.Phish.url. The first method involves phishers directing spam to end-users through their own machines with well designed software. By targeting users in this manner, phishers target big brands. The second method spammers target unpopular brands and their users through Botnet. Mostly the target contains a login and password page where users are asked to enter the login and password which, ultimately, are controlled by phishers. Their subject line is not properly designed like the previous one.

## 4.2 SPAM.HOAX_PHISH_FORGED_ CITIBNK
This phishing spam targets Citibank account holders to divulge their Citibank ATM/Debit card and PIN numbers. The link given in the mail takes the user to a non-secure site controlled by the phishers. Mostly these sites are owned by fraudsters using fake names. When we checked their websites, they seemed to have disappeared by the take down approach. The phishing related to Citibank was mostly short lived and the number of scam our domain received in this category is less as shown in the Figure 2.

## 4.3 Spam.hoax.Hoax_Phish_forged_paypall
This is another kind of phishing scam that targets Paypal users. Spam using this method ask users to divulge their Paypal identity by using social engineering techniques that tell users to update their identity as earlier as possible before doing any financial online activity such as online purchases, money transfers through email, and payment to the commercial websites. The phishing mail asks users to update their details before using the Paypal account. The user will be taken to a non-secure website controlled by fraudsters.

## 4.4 Clm.HTML.Phishing.Pay-110
This type of phishing scam doesn't target any particular type of institution or brand; instead it targets naive users to disclose credit card numbers, bank account information and various other personal details. Authors of this type of mail pretend to be from a legitimate authority and demand the confirmation of personal data. The working principle is similar to other phishing scams. Some of this mail might contain attachments but opening the attachments is not harmful. The mail contents are modified based on the target and users. Clm.HTML.Phishing.Pay-110 is a Trojan horse that does not spread automatically like other DDoS attack causing spam or install any software or modify the destination systems registry.

## 4.5 Spam.Hoax.HOAX_PHISH_FORGED_EBAY
This kind of phishing scam targets eBay users. The number of scams received by users are more than other scams as shown in figure 2. The take down approach of this brand does not completely eliminate the scam but, as seen in the table, there are fluctuations in the number of mail received by the users. This follows the same methodology used by other brand spoofers like Paypal, Citibank etc.

## 4.6 Trojan.spy.html.bankfraud.od
This spam is a kind of Trojan virus used for phishing. This Trojan targets all kinds of users and brands similar to spam.Phish.url where all information such as passwords, credit card details, and account details can be stolen. Various versions of this scam is also received by end users as shown in the figure 2. The following figure shows the traffic of all viruses, worms and Trojans.

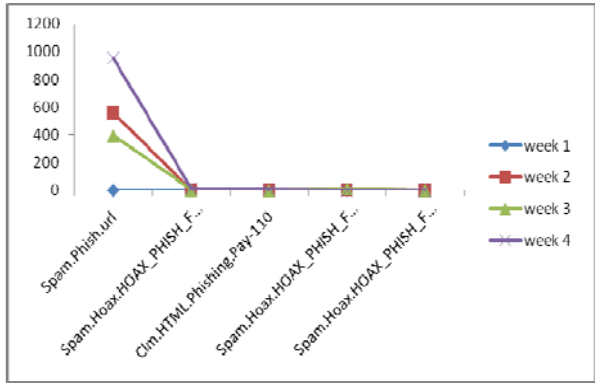

Figure 2. Virus, worm, Trojan traffic

**4.7 Password Phishing**

Apart from directly targeting financial institutions and its customers, phishers are targeting ordinary email users for possible financial gain by stealing email account.Fraudsters send emails to recipient to supply account details, if not within a week users email account will be closed. Fraudsters send this mail as its coming from the email service provider. If the user supplies their password and other details, phishers will change the password and lock the user out of their online account. They can read users email and access other personal information through the email account. If they find the right personal information, they can commit identity theft which includes using user's credit card details or steal money from user's bank accounts. If they don't get any useful information, the phishers will send email to other email accounts in the address book to trick them. The fraudulent email contains information that the user is in trouble or sick and needs financial assistance. The fraudsters will ask the end users contacts to send money in order to help the end user from suffering as shown in the figure.3 below.

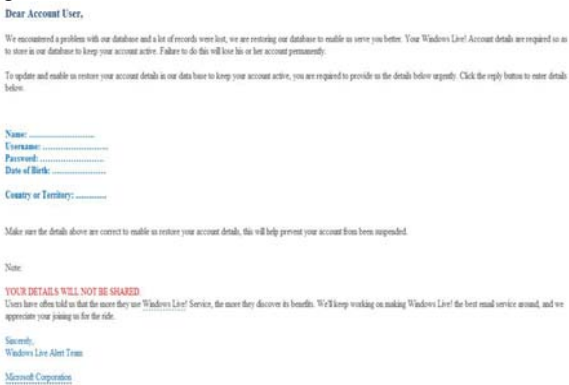

Figure 3. Sample Password phishing mail

If the above method doesn't work, phishers will use the users email account for spamming and phishing attacks Newly created email accounts have restrictions to send emails in a day, but this is not applicable for age old email accounts. So spammers target third party email accounts to spam end users and phishing.

## 5 Phishing Technology and Social Engineering

Phishers target end users with extraordinary precautions based on social engineering techniques. Instead of targeting the entire domain by using the brute force method, phishers send mail to a small group of email accounts (end users). On the first day, our domain received only 2 unique phishing mail for a particular brand. After that the number of email that target a particular brand and end users steadily increased as shown in Figure 4. After 4 to 5 days, once again, users got a reminder mail regarding their first phishing mail.

Phishers always send reminder mail to make the end users believe their contact is legitimate. If the phishers use brute force attack on the domain, the scam will reach the entire organization and the end users will be aware of the attack. To avoid this, phishers target small number of users first and increase the number of targets on the same domain day by day.

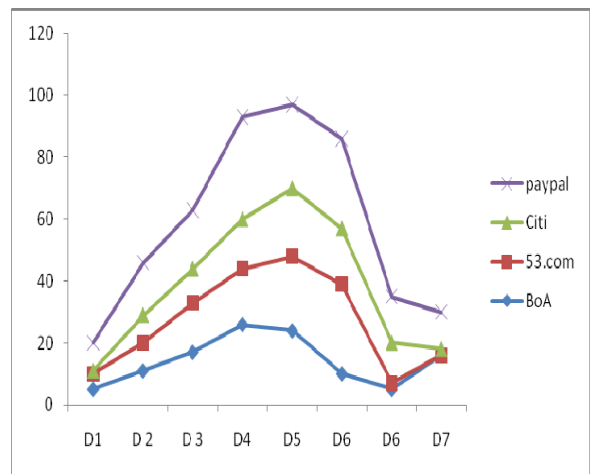

Figure 4.Phishing mails for particular target.

Phishers follow the end users regularly. After the first mail, they change their subject lines and contact the end user again. The sender's email account is used only once because a different email account is used to send reminder or confirmation mail. Usually the confirmation mail address contains the word refid (reference id) just to make the users believe it is a legitimate reminder.

**5.1 Methodology behind the attack:**
**Characterizing the sender ("from") address**

Phisher pays a lot of attention to social engineering techniques to lure end users. The design of the mail such as sender id, subject line, and contexts are carefully designed to lure victims.  These scam messages are sent by spam software which have the tools needed to generate email accounts as instructed. It can generate fake email accounts, subjects, duration of the spamming, domain name creation and modify the body of the message sent to the users [7][19]. This spam software can also generate the sender

email accounts with a specific format. The sender's account plays an important role to lure the victims to divulge the sensitive information. The sender's id designed such as real email accounts from the spoofed brands. The senders email account is designed exactly similar to the format of the legitimate user accounts. The length of the sender account is always more than 21 characters and up to 28 characters. It has three parts before the "at" symbol. The format of the sender's mail account is as follows.

(word1) (numericvalue) (word2)@ forged domain.com

Word1 contains sensitive words such as customerservice, support, operator, service-number, operator_id, Clientservice,ref, referencenumber, customers, accsupport. Phishers are using people who hold responsible jobs and have access to customer's information. The length of the second numeric part, numericvalue, ranges from 5 to 12 digits. In some cases word1 and numericword are connected by a "–" or "_" symbol as shown in the Figure 5. The length of word2, the third portion, is typically only 2 characters; mostly they are characters such as "ib", "ver", "ct" etc. as shown in the Figure.

| |
|---|
| customerservice-num903595453175ib@53.com |
| accsupport8013401973226ib@53.com |
| refnumber_id119700216971ver@security.53.com |
| reference_25676ver@security.53.com |
| referencenumber-7554431719494ct@citi.com |
| Referencenumber-843319866ib@citi.com |
| service_72673ib@citi.com |
| customers-20113070889ct@citi.com |

Figure 5. Sender email account syntax

**5.2 Analyzing the "subject" of the phishing mail**

Another important way to make user open spam is by issuing an attractive subject line, where the subject is carefully designed to lure victims. Examples of attractive subject lines include "to account confirmation", "message from the bank", "security warning", "update details", etc. By looking at the subject line, end users think incoming mail is legitimate and requires reading. Some of the subject lines also contain the date and time stamping to make end users believe the messages are legitimate. The format can include time of day, date month year, time (hour:minute:seconds-time zone). Often, the subject line contains only three time zones –"0800", "-0600", "-0500", all time zones from the US or Canada. Since all brands are North American users tend to believe the North American time zones. There is no other time zone stamping in phishing mails.

| |
|---|
| Fifth Third Bank - confirm your information! |
| Fifth Third Bank - secure confirmation |
| Bank of America - official information! |
| Citibank - Please Confirm Your Information |
| e-banking account confirmation |
| attention from fifth bank |
| security maintenance. |
| reminder: please update your details |

Figure 6. Example subjects

The phishing mail subject line can also confirm a bank name and phrases like "customer service" etc., with date and time to make the end-users believe that the message is legitimate. The end users get phishing mail every day of the week. Mostly phishing mail sent during weekends has the date and time in the subject area.

| |
|---|
| notification from Citibank. -Tue, 13 Jan 2009 08:15:07 -0800 |
| Fifth Third Bank reminder: confirm your account details -Tue, 13 Jan 2009 16:24:33 -0000 |
| Fifth Third Bank - urgent security notification for client. -Tue, 13 Jan 2009 08:31:15 -0800 |
| customer service: your account in Fifth Third Bank. –Tue, 13 Jan 2009 18:13:37 -0600 |
| Alert! [Tue, 13 Jan 2009 11:49:19 -0500] |

Figure 7. Example subjects with time stampings

**5.3 Difference between Phishing and DDoS attacks**

Phishing scams are mass-mailed by a group of criminals. Although received by e-mail, they do not spread themselves like DDoS attacks. DDoS attacks also spread through emails via phishing but DDoS targets the entire domain and causes a service interruption. Typically, phishing targets only financial institutions for financial gain. Phishing mail does not spread from users address book of the infected machines as DDoS attacks, which messes up registry settings and results in a system becoming inoperable. Where an infected user machines will download malicious code from the attacker's machine and damage the network of the victim, the phishing scam doesn't harm the victim's network.

# 6 Countermeasures: Multi Layer approach to defend Phishing attack

There are more than 100 anti-phishing tools freely available to combat phishing threats but none of these tools stop phishing effectively. Phishers always come up with new methodlogy to bypass anti-phishing countermeasuers. Like spamming, phishing is heavily dependent on social engineering techniques rather than technological innovations. From our study, we understand that most spammers involved in phishing activities follow the same methodlogies of Spammers. Effective anti-spam methods should be implemeted to countermeasures spam threats [4]. Since there is no bullet proof mechanism to combat spam, the combination of methods such as SURBL, DNSBL, rDNS etc and content filter techniques can help stop spam reach the end user's inbox. Defending against spam will help to reduce the risk of phishing attacks. User awareness also plays an important role to avoid being tricked by phishers. Sometimes, however, bostering more about the security will cause a negative impact on the user community [8]. Finally we conclude by recommending multi-layer spam. Phishing defending approach and user awareness are the best way to deal with Phishers so Phishers don't bring down user networks or cause any interruption; it just steals users information for finanacial gain.

Based on our analysis, we developed an anti-phishing methodology and implemented in our network. We monitored the phishing victims in our network for the last 24 months from February 2008 to February 2010 i.e. before implementing and after implementing the multi layer scheme. The results show that this approach is highly effective to prevent phishing attacks. The proposed approach reduced more than 80% of the false negatives and more than 95% of phishing attacks in our network.

## 6.1 Multilayer approach to defend Phishing attacks

In this paper we propose a multi layer approach to defend phishing attacks. This approach is a combination fine tuning of spamming techniques, periodically classifying false negative to spam, block the attackers IP addresses, using source and content filters to mark phishing attempt, educate and train users, report to service providers and target Institution. The results show that this approach is highly effective to prevent phishing attacks.

The multi layer approach is shown as figure 8. We implemented this approach in our mail system and monitored the results for a year from Feb 2009 to Feb 2010. We saw a significant improvement to avert phishing attacks in our community.

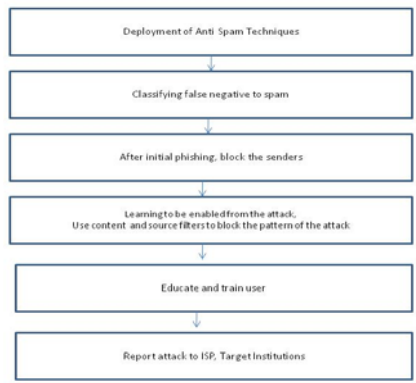

Figure.8 Proposed Multi layer approach to defend phishing attacks

The result shows that our approach is very effective. The approach has six layers as shown in figure 8. This approach is a combination of implementing anti spamming techniques, periodically classifying false negative to spam, after initial phishing attempt block IP addresses of source, enable source and content filters to mark phishing attempt, educate and train users and report to service provider and target Institution.

In the first step, we have to tighten anti spam techniques to counter all incoming phishing attempts through spamming. Even with strict anti spam techniques, some of spam will infiltrate user inbox. To avoid this, the network administrator should classify false negative to spam regularly. This will reduce the risk of phishing attacks. After initial phishing attacks in the domain, the email accounts, domains, IP address range of the sender should be blocked. In the next step, this attack method and pattern should be added to the database through learning. The content filters examine the contents of the mails and blocks the incoming unwanted mails. Network monitoring approach provides general solution to identify attacks prior and also during the attack. Institutions should educate user about possible phishing attack scenarios and how to act after being targeted. Sample phishing mails should be circulated to users to educate them. Institutions should advise users to report any possible phishing mails to network the administrator. If phishing attacks are identified, it should be reported to the service provider and target institutions. We implemented this approach for the last thirteen months from February 2009 to February 2010. The number of phishing victims, false negatives has gone down drastically as shown figure 9.

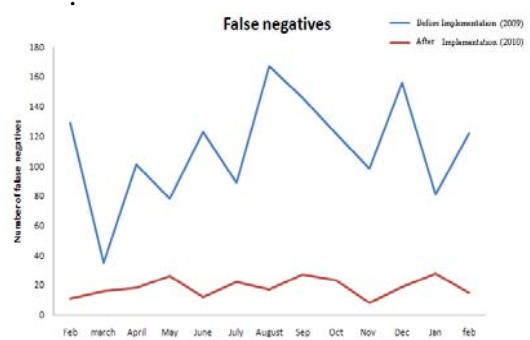

Figure 9. False negatives

We monitored the phishing attacks in our network for the last 24 months from February 2008 to February 2010 i.e before implementing and after implementing the proposed scheme. The above graph shows the number of false negatives before and after implementing the proposed mechanism. Before implementing this mechanism, the network received false negatives ranging from 35 to 167, at the average of 111 per month. But after implementing the proposed mechanism, the network users received false negatives ranging from 8 to 26 at the average of 18 per month. Our results show that the mechanism reduced more than 80% of false negative received by phishing attacks.

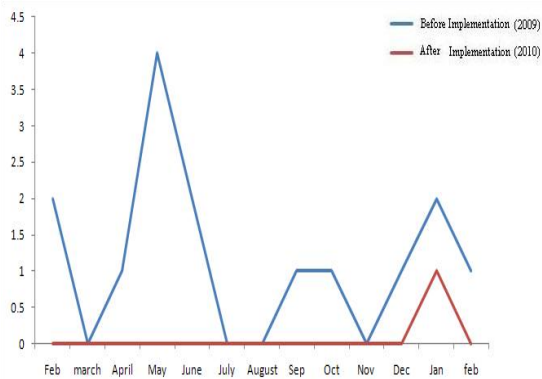

Figure 8. Phishing victims

We monitored the phishing victims in our network for the last 24 months from February 2008 to February 2010 i.e before implementing and after implementing the scheme. The above graph shows the number of victims before and after implementing the proposed mechanism. Before implementing this mechanism, the number of victims among the network users ranged from 0 to 4, at the average of 1.15 per month. But after implementing the proposed mechanism, the number of victims had gone down drastically as shown in figure 8. It shows that the mechanism reduced more than 99% of victims or incidents in the user community. The victim doesn't mean that he or she faced financial loss. Here the word "victim" for users that lost email accounts to the phishers and user that saved in lost moment of the financial loss. We have to mention that the users faced financial loss is less than .001% in our network or organization. But victims mentioned over here lost their personal email account due to the phishing attacks.

## 7 Conclusion

Phishing is a byproduct of spam and our data shows that a large number of victims of phishing are end users, brands and financial institutions. We analyzed millions of spam mail received in our server spam trap and our analysis shows two types of phishing attack. The first method used by phishers is to target end user email requesting they divulge sensitive informtion to the phishers controlled machines by clicking a link. The second method is through virus, malware and Trojan entities which infect and take users to fradulent websites hoping they divulge their idendity. Usually spamming techniques are used to reach an end user's inbox. Off late spammers are involved in phishing attacks. Phishers use social engineering techniques rather than software vulnerabilites and usually target users by sending mails to small group in the particular domain and then slowly increase their reach. After a few days, phishers start sending "reminder" mails to their initial mail targets to make them believe the original mail is legitimate. If they use the brute force method, phishers believe the user will think it's a hoax mail and prevent the user from accepting the mail. Spammers usually use a brute forec method only to spam end users and mimic DDoS attacks. Past studies show the number of spam mail goes down during the weekend, but our recent study [6] reveals that this is not true. Since spammers change their tactics and bypass spam countermeasuers, they don't minimize their activity during weekends. Since there are several specially designed software tools available to automatically generate spam, phishing can be performed on a continual basis. Based on our analysis, we designed an anti-phishing methodology and implemented in our network. The results show that the multi layer approach is highly effective to prevent phishing attacks. The proposed approach reduced more than 80% of the false negatives and more than 95% of phishing attacks in our network.